\documentclass{amsart}

\usepackage{fancyheadings}
\usepackage{amsfonts}
\usepackage{graphicx}
\usepackage{amscd,amsmath}
\usepackage{fancyhdr}
\usepackage{amssymb}
\usepackage{setspace}
 \newtheorem{thm}{Theorem}[section]
\newtheorem{eg}[thm]{Example}

\theoremstyle{remark}

\begin{document}
\large
\title[Multi-Prime RSA Over Galois Approach ]{Multi-Prime RSA Over Galois Approach }
\author [Swati Rawal ]{Swati Rawal}
\address{Department of Mathematics, Motilal Nehru National Institute of Technology Allahabad, Allahabad
(UP),India.}

\email{swati.rawal25@gmail.com}


\maketitle

Many variants of RSA cryptosystem exist in the literature. One of
them is RSA over polynomials based on Galois approach. In standard
RSA modulus of product of two large primes whereas in the Galois
approach author consider product of two irriduciable polynomials as
modulus. We use this idea and extend Multi-prime RSA over
polynomials. \vspace{.25cm}



\section {introduction}
RSA Cryptosystem is the first practical realization of the
public-key system invented by Rivest, Shamir and Aldeman
~\cite{RSA78}. One of the variant of RSA is obtainted by modifying
the RSA modulus i.e. Multiprime RSA ~\cite{Col97} and the other
variant is extension of RSA over polynomials ~\cite{Kar82}. We
extended the Multiprime RSA over polynomials .

\subsection{ RSA Cryptosystem} The best known public-key
cryptosystem is the RSA ,named after its inventors  Rivest,Shamir
and Adleman ~\cite{RSA78}. The RSA cryptosystem is defined as below:

\noindent\textbf{Key Generation:} We generate two randomly and large
primes $ p $ and $ q $ and computes the product \begin{center} $
n=pq $.
\end{center}
Then choose an integer $ e $ with $ 1 < e < \phi (n) $ and gcd
$ (e,\phi(n))=1 $.

Then compute the integer $ d $ with $ 1 < d < \phi (n) $ and $ de=1
mod \phi(n) $.

Since gcd $ (e, \phi (n))=1 $, such a number $ d $ exists. It can be computed by Extended Euclidean Algorithm.

Public key is pair $ (n,e) $ and Private key is $ (d,p,q) $. $ n $ is called the RSA modulus ,$ e $ is encryption exponent and $ d $ is decryption exponent.

\noindent\textbf{Encryption:} Let $ x \in Z_n $ be the plaintext
then $ x $ can be encrypted as
\begin{center}
$ e_k(x)=x^e modn $
\end{center}
$ e_k(x)=y $ is the ciphertext.

\noindent\textbf{Decryption:} If $ y \in Z_n $ is the ciphertext,
then $ x $ can be computed as
\begin{center}
$  x = d_k(y) = y^d modn $.
\end{center}

\subsection{ Multi-Prime RSA } As the name suggests in Multi-prime
RSA uses the modulus of the form $ N=p_1p_2...p_r $ product of more
than two primes introduced by Collins, Hopkins, Longford and Sabin
~\cite{Col97}. We first describe the key generation,encryption and
decryption as below:

\noindent \textbf{Key generation :}
The key generation algorithm takes as input a security parameter $ n $ and an
additional parameter $ b $. It generates an RSA public/private key pair as follows:
\begin{enumerate}
\item Generate $ b $ distinct primes $ p_1, . . . , p_b $ each $ \lfloor n/b \rfloor $-bits long. Set $ N  \longleftarrow \Pi^b_{i=1} p_i $. For a 1024-bit modulus we can use at most $ b = 3 (i.e., N = pqr) $.
\item Pick the same $ e $ used in standard RSA public keys, namely $ e = 65537 $. Then compute $ d = e^{-1} mod \phi(N) $ . As usual, we must ensure that $ e $ is relatively prime to $ \phi(N) = \Pi^b_{i=1} (p_i-1) $.
\end{enumerate}
The public key is $ (N, e) $; the private key is $ d $.

\noindent \textbf{Encryption :}
Given a public key $ (N, e) $, the encrypter encrypts exactly as in standard RSA.

\noindent \textbf{Decryption :}
Decryption is done using the Chinese Remainder Theorem (CRT). Let $ r_i = d mod (p_i-1) $. To decrypt a ciphertext $ C $ first compute,
  $ M_i = C^{r_i} mod p_i $  for each $ i, 1 \leq i \leq b $.\\
Then combines the $ M_i $'s using the CRT to obtain $ M = C^d mod N $.

\subsection{Extension of RSA Over Polynomials} This
cryptosystem was proposed by Karvitz and Reeds in 1982
~\cite{Kar82}. In standard RSA cryptosystem the modulus $n$ is the
product of two primes, and the security of RSA was depends on
factoring the modulus $n$. Karvitz and Reeds took two irreducible
polynomials and then considered the product of two polynomials as
the RSA modulus.  The key generation, encryption and decryption is
defined as below:

\noindent\textbf{Key generation :} Let $ F $ be a finite field and
choose two irreducibles polynomials $ p_1(x) $ and $ p_2(x) $ of
higher order of say $ n_1 $ and $ n_2 $ and then compute $ f(x) $ =
$ p_1(x)p_2(x) $ of degree $ n $ = $ n_1 + n_2 $.

Choose a $ d $ such that it is relatively prime to $ (|F|^{n_1} - 1)(|F|^{n_2} - 1) $ and then compute $ e $ such that $ ed $ = $ 1 \mod(|F|^{n_1}-1)(|F|^{n_2}-1) $.

Then public key will be $ e $ and the secret key will be $ d $.

\noindent \textbf{Encryption :} Let $ m $ be message, put the
message into $ |F| $-array representation and break it into block
each of size $ n $.

Each block is associated with a polynomial over $ F $ of degree less than $ n $.

The plaintext is encrypted by $ C(x) $  = $ (M(x)^e (mod f(x)) $,
then $ c(x) $ will be the ciphertext.

\noindent \textbf{Decryption :} If $ C(x) $ is the ciphertext then
using the decryption exponent $ d $ then the plaintext can be
computed as  $ M(x) = C(x)^d (mod f(x)) $ .

\section{Proposed Scheme Based On Galois Approach}
we will introduce a new scheme based on Galois approach by modifying
the RSA modulus as we have discussed in multiprime RSA. Before
considering that scheme we introduce the notion of Chinese remainder
theorem over polynomials.
\begin{thm}
\textbf{Chinese Remainder Theorem Over Polynomials :}\\
Suppose $ m_1(x),......,m_r(x) $ be polynomials that are relatively prime to each other.
Let $ a_1(x),....,a_n(x) $ be polynomials and then the following system of congurences :

\begin{center}
$  p(x) \equiv a_1(x) (mod m_1(x)) $ \\
$  p(x) \equiv a_2(x) (mod m_2(x)) $ \\
   . \\
   . \\
   . \\
$  p(x) \equiv a_r(x) (mod m_r(x)) $ \\
\end{center}
has a unique solution modulo $ M(x)=m_1(x)\times m_2(x)....m_r(x) $, which is given by \\ \begin{center}
$ \sum_{i=1}^{r} a_i(x)M_i(x)y_i(x) modM(x), $\end{center}
where $ M_i(x)=M(x)/m_i(x) $ and $ y_i(x)=M_i(x)^{-1} modm_i(x) $, for $ 1 \leq i \leq r(x) $ .\\
\end{thm}

Using the above stated theorem we can extent the concept of RSA-CRT over polynomials also.Let us consider the key generation ,encryption and decryption as following : \\

\noindent \textbf{Key Generation :} Same as RSA over Galois Approach.

\noindent \textbf{Encryption :} Same as we have discussed in RSA over Galois Approach.

\noindent \textbf{Decryption :}

If $ c(x) $ is the ciphertext and $ d $ is the private key then,
first Compute,
\begin{center}
$ d_p = d mod |F|^{n_1}-1 $   and   $ d_q = d mod |F|^{n_2}-1 $
\end{center}
then Compute
\begin{center}
 $ M_p(x) = C^{d_p} mod p_1(x) $  and  $ M_q(x) = C^{d_q} mod p_2(x) $
 \end{center}
Now, using Chinese remainder theorem we can find the plaintext $ m $ as ,
\begin{center}
$ M(x) = y_{p_1}(x)M_{p_1}(x)p_1(x) + y_{p_2}(x)M_{p_2}(x)p_2(x) $ , where
\end{center}
$ y_{p_1}(x)= M^{-1}_{p_1}(x) mod p_2(x) $ and $ y_{p_2}(x)= M^{-1}_{p_2}(x) mod p_1(x) $.

Now, We will consider the Multiprime RSA over polynomial as here we
will modify the RSA modulus by considering it of the type $ f(x) $ =
$ p_1(x)p_2(x)...p_b(x) $, product of more than two irreducible
polynomials of large degree.

Let us consider its key generation , encryption and decryption :

\noindent \textbf{Key generation : }
The key generation algorithm takes as input a security parameter $ n $ and an
additional parameter $ b $. It generates an RSA public/private key pair as follows:
\begin{enumerate}
\item Generate $ b $ distinct irreducible polynomials $ p_1(x), . . . , p_b(x) $ each of degree $  n/b  $ over the finite field $ F $. Set $ f(x) = \Pi^b_{i=1} p_i(x) $.

\item Choose a $ d $ such that it is relatively prime to $ \phi_F(f(x) $ = $ (|F|^{n_1}-1)(|F|^{n_2}-1),...(|F|^{n_b}-1) $ and compute $ e $ such that $ e $ = $ d^{-1} mod \phi_F(f(x)). $
\end{enumerate}
The public key is $ (f(x), e) $; the private key is $ d $.

\noindent \textbf{Encryption :} Same as Standard RSA over polynomials as we have discussed before.

\noindent \textbf{Decryption :}
Decryption is done using the Chinese Remainder Theorem (CRT). Let $ r_i = d mod (|F|^{n_i}-1) $. To decrypt a ciphertext $ C(x) $, first computes

 $ M_i(x) = C(x)^{r_i} \mod p_i(x) $  for each $ i, 1 \leq i \leq b $.\\
Then combines the $ M_i(x) $'s using the CRT to obtain $ M(x) = C(x)^d mod N $. \\

The advantage of proposed scheme over ~\cite{Col97} is same as the
advantage of multyprime RSA ~\cite{Col97} and RSA with CRT
~\cite{Qui82} over standard RSA ~\cite{RSA78}.
\begin{eg}
Let us consider an example of above scheme we use $ b=3 $,
Consider the three irreducible polynomials $ p_1(x) $ = $ x^3+x+1 $ , $ p_2(x) $ = $ x^3+x^2+1  $ and $ x^2+x+1 $ over $ Z_2 $.

we compute $ f(x) $ = $ p_1(x)*p_2(x)*p_3(x) $ = $ x^8+x^6+x^5+x^4+x^3+x^2+1 $\\
then $ \phi(f(x)) $ = $ (2^3-1)(2^3-1)(2^2-1) $ = $ 147 $.\\
We choose $ e $ = 34 randomly such that $ gcd(e,\phi(f(x)) $ = $ 1 $, now compute $ d $ = $ e^{-1} $ mod $ \phi(f(x)) $ = $ 34^{-1} mod 147 $ = 13.

let $ m(x) $ = $ x^4+x^3+1 $ be the plaintext the on encryption we get,

\begin{center}
 $ c(x) $ = $ x^4+x^3+1^{34} $ mod$ f(x) $ = $ x^5+x^4+x^3+x^2+1 $.
 \end{center}

and now for decryption, compute
$ r_1 $ = $ d mod(|F|^{n_1}-1) $ = 6 , $ r_2 $ = $ d mod(|F|^{n_2}-1) $ = 6 and $ r_3 $ = $ d mod(|F|^{n_3}-1) $ = 6.
then,

$ m(x) $ = $ c(x)^{r_1} $ mod$ p_1(x)  $ = $ x+1 $ \\
$ m(x) $ = $ c(x)^{r_2} $ mod$ p_2(x)  $ = $ x^2 $ \\
$ m(x) $ = $ c(x)^{r_3} $ mod$ p_3(x)  $ = $ 1 $ \\ then using chinese remainder theorem plaintext will be

\begin{center}
$ m(x) $ = $ c(x)^d modf(x) $ = $ x^4+x^3+1 $
\end{center}
\end{eg}
The above example can be easily implemented using MATLAB to avoid long calculations.


\normalsize
\begin{thebibliography}{}
\bibitem{Col97} Collins T., D. Hopkins, Longford S. and Sabin M., Public key cryptography operates and method. US. Patent 5,848, 159, Jan. 1997.
\bibitem{Kar82} Karvitz D.W., Reeds I.S., An Extension of cryptostructure: An Galois Approach ,Electronic Letters 18 March 1982,v.18 n.6,255-256.
\bibitem{RSA78}R. Rivest, A. Shamir and L. Adleman ;  A Method for Obtaining Digital Signatures and Public-key Cryptosystem;  Communications of the ACM, February, 1978.
\bibitem{Qui82} Quisquater and Couvruur, Fast deciphering algorithm for RSA public key cryptosystem. Electronics Letters v. 01, 18, 1982, 905-907 .

\end{thebibliography}
\end{document}